\documentclass[12pt,aps,prb,reprint,amsmath,amssymb,bibnotes,longbibliography]{revtex4-1}
\usepackage{graphicx,makecell} 

\bibpunct{}{}{,}{s}{}{}
\makeatletter \renewcommand\@biblabel[1]{$^{#1}$} \makeatother

\begin{document}

\title{A Spectral Analysis of Laser Induced Fluorescence of Iodine}

\author{S. B. Bayram}\email{bayramsb@miamioh.edu}
\affiliation{Miami University, Physics Department, Oxford, OH 45056
}
\author{M. V. Freamat}\email{freamamv@morrisville.edu}
\affiliation{Morrisville State College, Physics Department, Morrisville, NY 13408
}
\date{\today}

\begin{abstract}
When optically excited, iodine absorbs in the 490- to 650-nm visible region of the spectrum and, after radiative relaxation, it displays an emission spectrum of discrete vibrational bands at moderate resolution. This makes laser-induced fluorescence spectrum of molecular iodine especially suitable to study the energy structure of homonuclear diatomic molecules at room temperature. In this spirit, we present a rather straightforward and inexpensive experimental setup and the associated spectral analysis which provides an excellent exercise of applied quantum mechanics fit for advanced laboratory courses. The students would be required to assign spectral lines, fill a Deslandres table, process the data to estimate the harmonic and anharmonic characteristics of the ground vibronic state involved in the radiative transitions, and thenceforth calculate a set of molecular constants and discuss a model of molecular vibrator.
\end{abstract}

\pacs{0150Pa @ Laboratory experiments and apparatus} \maketitle

\section{\label{sec:introduction}Introduction}
Courses of atomic and molecular spectroscopy are valuable components of advanced physics curricula not in the least because they provide a rich ground for quantum mechanical applications. For example, in the advanced lab course of spectroscopy offered in the Physics Department at Miami University, \cite{Sand07, Bayr09, Blue10, Bayr12a, Bayr12b} the students perform a sequence of experiments exposing various facets of the inner clockwork of atomic and molecular structures. 
In this paper we introduce one of these instructional experiments: the spectral analysis of laser induced fluorescence of molecular iodine I$_2$ at room temperature. The educational value of this experiment is two-fold: On one hand it helps students practice some indispensable spectrometry skills inasmuch as they assign electron transitions based on selection rules and Franck-Condon Principle, or an iodine atlas. On the other hand, they learn how to implement a quantum mechanical model for the diatomic molecule as they use the solution of Schr\"{o}dinger equation with a Morse potential to find the transition energies in terms of the fundamental vibrational frequency of the molecule and the first order deviation from a harmonic oscillator. This spectrometric study complements two other experiments treating the vibrational and rotational spectrometry of the diatomic molecule of nitrogen.\cite{Bayr12b,Bayr15}

\section{\label{sec:background}Theoretical background}
\subsection{Iodine absorption and emission}
Iodine is the heaviest common halogen and, like the other halogens, in gas phase forms weakly bound diatomic molecules. Iodine has an unusually long bond length and its vapor absorbs light in the visible spectrum, in the yellow region, such that the gas appears violet. The molecular orbitals are characterized by strong spin-orbit coupling, such that the degeneracy of the state associated with the electronic angular momentum of the excited molecule is lifted, which results in rather complex absorption patterns in the visible.\cite{McHa99}
Therefore, iodine has a richer configuration of non-degenerate excited states than lighter homonuclear diatomic molecules. Within each of these electronic configuratios, there is a fine structure of vibrational states, which in turn contain rotational levels,\cite{Herz50} such that the total energy can be partitioned as
\begin{equation}
\label{eq:totEnergy}
E = E_{el} + E_{vib}+E_{rot}.
\end{equation}

\begin{figure}[!ht]
\centering
\includegraphics[scale=0.7]{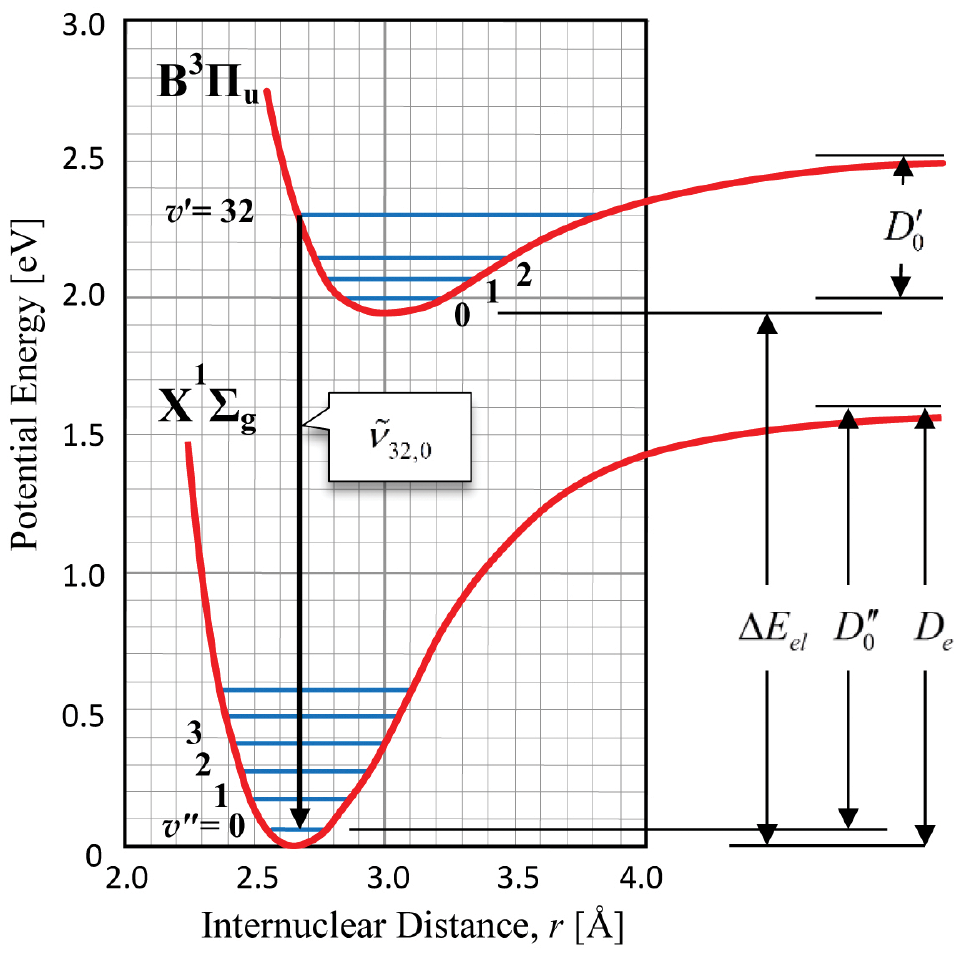} 
\caption{\label{fig:Potentials} Schematic potential curves for the $X$ and $B$ states involved in the observed fluorescence of iodine.\cite{Mull71} Note the transition responsible for the spectral maximum. }
\end{figure}

In our experiment we are interested in the absorption followed by radiative emission between the vibrational levels of the singlet ground state $X^1\Sigma_g^+$ and the first non-dissociative excited triplet state $B^3\Pi_u^+$. A detailed description of the selection rules governing electronic transitions and the respective spectral notation are given in Ref.[\citenum{Mull30}]. It is interesting to note that the maxima of the absorption and ensuing emission spectra are shifted relative to each other, and with respect to the transition from the lowest vibrational state of the excited level. To understand why, note that at room temperature the electrons populate mostly the lowest vibrational levels of the ground state, indexed by the quantum number $v''$. Under incident visible light, the molecule will absorb photons and undergo vibronic transitions to various vibrational levels indexed $v'$ in the excited state. Subsequently, the molecule will relax in two steps, first to a single vibrational level in the excited state and then back to a range of low-lying vibrational levels in the ground state, such that the maximum of the emission spectrum will occur at a higher wavelength then the absorption maximum. On the other hand, to explain the shift from the $v'=0 \to v''$ transition, one needs to use the Franck-Condon Principle, which states that the vibronic transitions occur almost instantaneously compared to the slow response of the atomic nuclei shifting to other inter-atomic separation. The idea can be clarified using the potential curves and the respective structure of vibrational levels for the ground and excited states, as illustrated for iodine in Fig.~\ref{fig:Potentials}. Note from the figure that the electronic reconfiguration resulting from vibronic transitions will reshape the molecule and thence the bond strength and length. According to the principle, the transitions will be essentially vertical, driven by the corresponding overlap between the wavefunctions in the ground and excited states, and thence strongly influenced by the relative offset of the two potential curves. Consequently, due to the significant difference in bond length, the $B \to X$ transitions will occur only from an upper vibrational state $v'$ preponderantly to the lowest levels $v''$.

\subsection{Quantum mechanical model}
Traditionally, the spectral activity of iodine is mostly studied for instructional purposes using its absorption spectrum, which can be used to characterize the excited state $B$.\cite{Lewi94} However, in our experiment the students  stimulate the iodine vapor with a laser and analyze the emission spectrum to extract some properties of the ground state $X$. The approach is fairly simple, but instructive.
\begin{figure*}[!ht]
\centering
\includegraphics[scale=0.9]{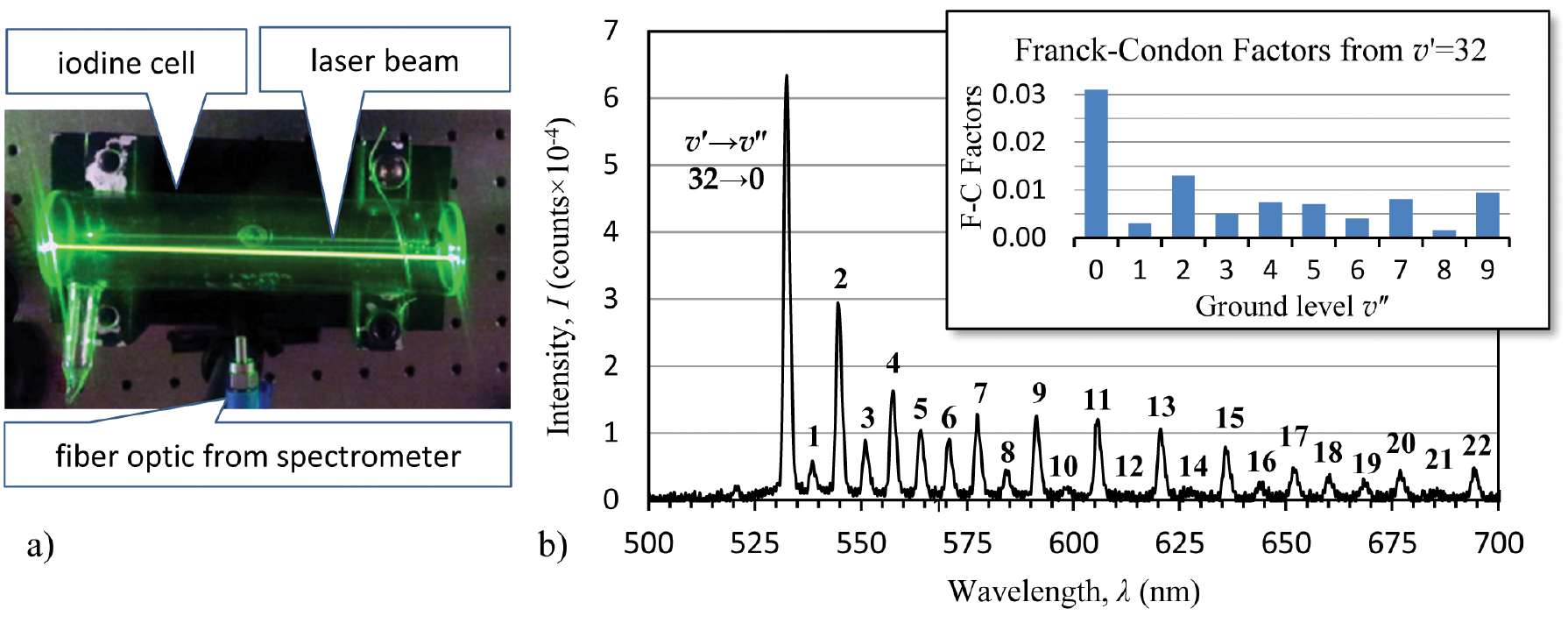} 
\caption{\label{fig:SetupSpectrum} a) Experimental setup to record laser-induced fluorescence using a green laser diode at 532 nm, an iodine cell, and a spectrometer. b) The $B\to X$ emission spectrum. The inset shows the Franck-Condon factors\cite{Tell07} for the $v'=32 \to v''$ transitions at 532 nm.}
\end{figure*}
To adapt the formalism to the standard spectrometric units, it is customary to define energies in terms of wavenumbers in units of cm$^{-1}$ by dividing Eq.~(\ref{eq:totEnergy}) by $hc$. The resulting quantity is called a term value $T$. Thus, ignoring the rotational energy levels which cannot be resolved at the resolution of our experiment, the term value of a vibronic state becomes $T=E/hc = T_{el} + G_v$, where $G_v=E_{vib}/hc$ is the standard notation for the term value of a vibrational level of quantum number $v$. With this rescaling, the frequency for a transition between two vibronic states $v' \to v''$ is given by
\begin{equation}
\label{eq:transFreq}
\tilde{\nu}_{v',v''}= T' - T''= \omega_{el} + G_{v'}-G_{v''},
\end{equation}
where $\omega_{el}= T'_{el} - T''_{el}$ is the frequency of the electronic transition.
An approximated form for the vibrational term value can be obtained by solving Schr\"{o}dinger equation without the rotational term, in combination with a well selected model potential, such as the asymmetric Morse potential:\cite{Mors29}
\begin{equation}
\label{eq:Morse}
V(r)=D_e\left(1-e^{-\beta(r-r_0)}\right)^2,
\end{equation}
where $D_e$ is the dissociation energy given by the depth of the potential well, and $\beta$ is related to the force constant $k$ of the molecule:
\begin{equation}
\label{eq:beta}
\beta=(k/2D_e)^{1/2}.
\end{equation}
With this potential, the vibrational term value of each level $v$ can be written:
\begin{equation}
\label{eq:valTerm}
G_v =\omega_0(v+\tfrac{1}{2})-\omega_0x_0(v+\tfrac{1}{2})^2,
\end{equation}
where $\omega_0$ is the fundamental frequency, and the \textit{anharmonicity} product $\omega_0x_0$ models the first order deviation from a harmonic oscillator.  The fundamental frequency of the vibration, $\omega_0$, is naturally related to the other parameters of the oscillator:
\begin{equation}
\label{eq:harmFreq}
\omega_0 = \frac{1}{2 \pi c}\sqrt{\frac{k}{\mu}}=\frac{\beta}{c}\sqrt{\frac{D_e}{2\pi^2\mu}},
\end{equation}
where $\mu = 1.05\times10^{-25}$ kg is the reduced mass of the iodine molecule.
 \subsection{Molecule characterization}
The term value solution in Eq.~(\ref{eq:valTerm}) can be used to extract information about the ground potential. One of the most popular method is the Birge-Sponer treatment.\cite{Spon26} This approach requires the subtraction of energies for transitions originating from the same excited vibrational level into ground states of successive vibrational indices $v''$: $\Delta \tilde{\nu}_{v''} = \tilde{\nu}_{v',v''}-\tilde{\nu}_{v',v''+1}$. Then, Eqs.~(\ref{eq:transFreq}) and (\ref{eq:valTerm}) yield
\begin{equation}
\label{eq:DeltaE}
\Delta \tilde{\nu}_{v''} =G_{v''+1}-G_{v''}= \omega_0-2\omega_0x_0\left(v''+1\right).
\end{equation}
This expression indicates that the energy spacings are fairly constant at low $v''$, but reduce gradually at high $v''$, up to a maximum level $v''_{max}$ corresponding to the dissociation energy $D_e$. Based on this argument, the students plot $\Delta \tilde{\nu}_{v''}=\Delta G_{v''}$ versus $(v''+1)$ and fit it with a straight line. Then, the fundamental frequency $\omega_0$ is the intercept of the $\Delta G_{v''}$ axis, and the anharmonic term $\omega_0 x_0$ comes from the slope of the line. The dissociation energy $D_0$ with respect to the zero-point level is simply given by the area under the linear plot in the interval $(0,v''_{max})$. Because $\Delta \tilde{\nu}_0\approx \omega_0$ whereas $\Delta \tilde{\nu}_{v_{max}}\to0$, the zero-point dissociation energy
\begin{equation}
\label{eq:D0}
 D_0/hc = \omega_0^2/4\omega_0x_0.
\end{equation}
Combining with Eq.~(\ref{eq:valTerm}), the students can furthermore calculate the dissociation energy $D_e$ relative to the equilibrium point (that is, the bottom of the potential well) as shown in Fig.~\ref{fig:Potentials}:
\begin{equation}
\label{eq:De}
\frac{D_e}{hc} =G_0 +\frac{D_0}{hc} = \frac{1}{2}\omega_0-\frac{1}{4}\omega_0x_0 + \frac{\omega_0^2}{4\omega_0x_0}.
\end{equation}

\section{\label{sec:data}Experimental Setup and Data}
A straightforward and effective experimental arrangement can be used to produce an emission spectrum by inducing fluorescence in a iodine vapor cell using an inexpensive laser diode (Fig.~\ref{fig:SetupSpectrum}a). The 532-nm laser excites the $v'=32$ vibrational level of the excited $B$ state of iodine. The iodine gas in the cell needs to be pure enough to avoid quenching of the fluorescence by collisional dissociation of the iodine in the B-state.\cite{Tell07} In our lab, the spectrum is collected using an Ocean Optics-UV-VIS spectrometer with a resolution of 0.5 nm, sufficient to resolve the vibrational features. The spectrum is calibrated using a standard light source, such as a low-pressure mercury discharge tube which exhibits a bright green at 546.1 nm and two yellow lines at 579.0 nm and 577.1 nm.\cite{NIST}

As shown in Fig.~\ref{fig:SetupSpectrum}(b), the spectrum exhibits a maximum near 532 nm corresponding to the resonant fluorescence and a progression of regularly spaced lines with a width indicating the underlying rotational structure. The figure includes the Franck-Condon factors (FCF) which help assigning the vibrational quantum numbers for the observed peaks. It is readily noticeable that the pattern of spectral intensities match the FCF corresponding to the lowest excited vibrational level populated at 532 nm, $v'=32$, such that the spectral lines likely originate from this level in the $B$-state as the molecule relaxes to the ground vibrational levels, $v''=0,1,2...$. Also, the small but observable intensity for levels where it should be negligible (such as $v''=10$) indicates  a certain contribution from transitions originating in excited levels other than $v''=32$. The lines assigned on the figure are used for the sample analysis in this work.
Prior to processing the data, students can inspect the spectrum and discuss some of its characteristics. For instance, note the increasing separation between spectral lines at higher wavelength which indicates that the anharmonicity of the vibrational states increases with $v''$.

\begin{figure}[!ht]
\centering
\includegraphics[scale=0.85]{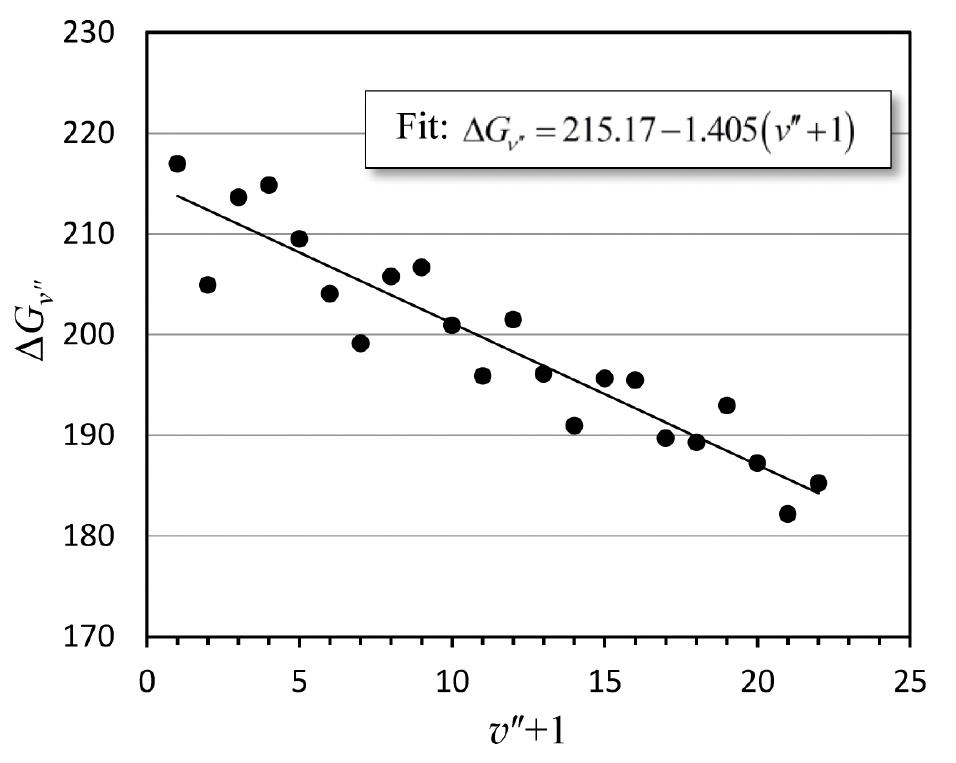} 
\caption{\label{fig:BirgeSponer} The Birge-Sponer method applied to the spectrum in Fig.~\ref{fig:SetupSpectrum}(b). 
}
\end{figure}

\section{\label{sec:analysis}Data Analysis}
Based on the spectrum, the students tabulate the data in a Deslandres table and then proceed with the treatment prescribed by the Birge-Sponer method by computing the differences $\Delta \tilde{\nu}_{v''} =\Delta G_{v''} =G_{v''+1}-G_{v''}$ and plotting the $\Delta G_{v''}$ versus $v''+1$ through as many ground levels as deem necessary. The unresolved rotational structure will impact the linearity of the plot mostly in the vicinity of the bandhead, but the linear fit will result in parameters fairly close to the expected values. As described above, the $\Delta G_{v''}$ intersect will yield the harmonic frequency $\omega_0$, whereas the anharmonic term  $\omega_0 x_0$ is half the slope. Based on these parameters, Eqs. (\ref{eq:D0}) and (\ref{eq:De}) can be employed to calculate the dissociation energies, and thenceforth Eqs. (\ref{eq:harmFreq}) and (\ref{eq:beta}) to characterize the Morse potential as a model for the ground state $X$.
For example, when treated by this method, the sample spectrum in Fig.~\ref{fig:SetupSpectrum}(b) provides the data for the Birge-Sponer plot in Fig.~\ref{fig:BirgeSponer}, which in turn generates the parameters tabulated in Table~\ref{tab:Results}. These results are compared to accepted values to emphasize how, albeit simple both in terms of experimental setup and analytical approach, the experiment delivers estimates of the fundamental frequency and anharmonic term remarkable close to the expected value. The dissociation energy is larger than the accepted value likely because the resolution of the spectrometer is insufficient to resolve the fine rotational structures, such that assigning the progression is affected by errors which, albeit small, may alter the slope of the fit line.
\begin{table}[!ht]
\centering
\begin{tabular}{|c|*{4}{c}|}\hline
\makecell*[c]~Constants~& $\omega_0$ [cm$^{-1}$] & $\omega_0x_0$ [cm$^{-1}$] & $D_e$ [eV] & $k$ [N/m]  \\
\hline
\makecell*[c]~Experiment~& 215 & $0.70$  & 2.0& 173 \\
\makecell*[c]~Literature~& 214 & 0.67 & 1.54 & 172 \\
\hline
\end{tabular}
\caption{The molecular constants for the ground state $X$ of iodine compared to values published in literature.\cite{McNa80}}
\label{tab:Results}
\end{table}

\section{\label{sec:Conc}conclusions}
The iodine molecules are excited at 532 nm from the ground state $X$ into the $v'=32$ vibrational level of the $B$ state. The fluorescence spectrum from the $B$ $\to$ $X$ states, exhibits a progression of distinct lines in the vicinity of the laser wavelength, can be readily processed by the students to find a set of molecular constants. The experiment provides a useful exercise in applied quantum mechanics as it requires the students to conceptualize the iodine fluorescence in the light of the quantum molecular theory, employ a simple asymmetric potential to make predictions about the quantized transition energies, and apply the model to the data set using the Birge-Sponer method to extract the harmonic and anharmonic parameters of the ground potential, as well as the dissociation energy and force constant of the potential used to emulate the structure of vibrational levels.

\begin{acknowledgments}
Financial support from the National Science Foundation
(Grant No. NSF-PHY-1309571) and Miami University, College of Arts and Science is gratefully acknowledged.
\end{acknowledgments}



\begin{thebibliography}{26}
\bibitem{Sand07}B. L. Sands, M. J. Welsh, S. Kin, R. Marhatta, J. D. Hinkle, and S. B. Bayram,
 Am.~J.~Phys. {\bf 75}, 488-495 (2007).
\bibitem{Bayr09} S. B. Bayram, ``Vibrational spectra of the nitrogen molecules in the liquid and gas phase,'' presented at the AAPT Topical Conference on Advanced Laboratories, Ann Arbor, MI, 2009.
\bibitem{Blue10} J. Blue, S. B. Bayram, and S. D. Marcum,
Am.~J.~Phys. {\bf 78}, 503-509 (2010).
\bibitem{Bayr12a} S. B. Bayram and M. V. Freamat, ``The Advanced Spectroscopy Laboratory Course at Miami University,'' presented at the AAPT/ALPhA Conference on Laboratory Instruction Beyond the First Year of College, Philadelphia, PA, 2012.
\bibitem{Bayr12b} S. B. Bayram and M. V. Freamat,
Am.~J.~Phys. {\bf 80}, 664-669 (2012)
\bibitem{Bayr15} S. B. Bayram and M. V. Freamat,
accepted for publication in Am.~J.~Phys., arXiv:1504.05510 [physics.ed-ph]
\bibitem{McHa99} Jeanne L. McHale, \textsl{ Molecular Spectroscopy }, 3rd. ed. (Prentice Hall, 1999), pp.319.
\bibitem{Herz50} G. Herzberg, \textsl{Molecular Spectra and Molecular Structure I. Spectra of Diatomic Molecules}, 2nd. ed. (Van Nostrand, Princeton, 1950).
\bibitem{Mull30} R. S. Mulliken,
Rev. Mod. Phys.~{\bf 2}, 60--115 (1930).
\bibitem{Mull71} R. S. Mulliken,
J. Chem. Phys., {\bf55}, 288-309 (1971).
\bibitem{Lewi94} E. L. Lewis, C. W. P. Palmer, and J. L. Cruickshank,
Am.~J.~Phys. {\bf 62}, 350-356 (1994).
\bibitem{Mors29} P. M. Morse,
Phys. Rev. {\bf 34}, 57-74 (1929).
\bibitem{Spon26} H. Sponer and R. T. Birge,
Phys. Rev. {\bf 28}, 256 (1926).
\bibitem{Tell07} J. Tellinghuisen,
J. Chem. Educ. {\bf 84}, 336-341 (2007).
\bibitem{McNa80} I. J. McNaught,
 J. Chem. Educ. {\bf 57}, 101-105 (1980).
\bibitem{NIST} NIST, physics.nist.gov/PhysRefData/Handbook/Tables.
\end{thebibliography}
\end{document}